\RequirePackage{amsmath}

\documentclass[runningheads]{llncs}
\usepackage[T1]{fontenc}
\usepackage{graphicx}
\usepackage{float}

\usepackage[ruled, vlined, linesnumbered]{algorithm2e}
\usepackage{hyperref}
\usepackage{multirow}
\usepackage{makecell}
\usepackage{multicol}
\usepackage{booktabs}
\usepackage{amssymb}
\usepackage{amsmath}
\floatstyle{plaintop}
\usepackage{subcaption}

\usepackage[newfloat,frozencache,cachedir=.]{minted}
\newenvironment{code}{\floatstyle{plaintop}%
\captionsetup{type=listing, labelfont=bf,justification=raggedright, singlelinecheck=false}}{}
\SetupFloatingEnvironment{listing}{name=\textbf{Listing}}

\makeatletter
\newcommand*\mysize{%
  \@setfontsize\mysize{8}{8}%
}
\makeatother
\setminted{fontsize=\mysize}

\usepackage{enumitem}

\newcommand{\benchmark}[0]{\textsc{C-Pack-IPAs}\xspace}
\newcommand{\cfaults}[0]{\textsc{CFaults}\xspace}
\newcommand{\bugassist}[0]{\textsc{BugAssist}\xspace}
\newcommand{\sniper}[0]{\textsc{SNIPER}\xspace}
\newcommand{\tcas}[0]{\textsc{TCAS}\xspace}
\newcommand{\cbmc}[0]{\textsc{CBMC}\xspace}

\usepackage[misc]{ifsym}
 
\begin{document}
\title{\cfaults: Model-Based Diagnosis for Fault Localization in C with Multiple Test Cases}
\titlerunning{Model-Based Diagnosis for Fault Localization in C with Multiple Test Cases}
%
\author{Pedro Orvalho\inst{1}\orcidID{0000-0002-7407-5967}(\Letter) \and
Mikoláš Janota\inst{2}\orcidID{0000-0003-3487-784X} \and
Vasco Manquinho\inst{1}\orcidID{0000-0002-4205-2189}}
\authorrunning{P. Orvalho et al.}
%
\institute{INESC-ID, IST, Universidade de Lisboa, Portugal\\
\email{\{pmorvalho, vasco.manquinho\}@tecnico.ulisboa.pt}\\ 
\and
Czech Technical University in Prague, Czechia\\\email{mikolas.janota@cvut.cz}}
\maketitle              
\begin{abstract}

Debugging is one of the most time-consuming and expensive tasks in software development. Several formula-based fault localization (FBFL) methods have been proposed, but they fail to guarantee a set of diagnoses across all failing tests or may produce redundant diagnoses that are not subset-minimal, particularly for programs with multiple faults.

This paper introduces a novel fault localization approach for C programs with multiple faults. \cfaults leverages Model-Based Diagnosis (MBD) with multiple observations and aggregates all failing test cases into a unified MaxSAT formula. Consequently, our method guarantees consistency across observations and simplifies the fault localization procedure. 
Experimental results on two benchmark sets of C programs, \tcas and \benchmark, show that \cfaults is faster than other FBFL approaches like \bugassist and \sniper. Moreover, \cfaults only generates subset-minimal diagnoses of faulty statements, whereas the other approaches tend to enumerate redundant diagnoses.

\end{abstract}
\section{Introduction}
\label{sec:intro}

Localizing system faults has always been one of the most time-consuming and expensive tasks. Given a buggy program, \emph{fault localization (FL)} involves identifying locations in the program that could cause a faulty behaviour (bug). 

Given a faulty program and a test suite with failing test cases, current \emph{formula-based fault localization (FBFL)} methods encode the localization problem into several optimization problems to identify a minimal set of faulty statements (diagnoses) within a program. 
Typically, these methods find a minimal diagnosis considering each failing test case individually rather than simultaneously with all failing test cases. Moreover, these FBFL methods enumerate all \emph{Minimal Correction Subsets (MCSes)}~\cite{DBLP:journals/jar/LiffitonS08} to cover all diagnoses.

For instance, \bugassist~\cite{bugAssist-cav11,bugAssist-pldi11}, a prominent FBFL tool, implements a ranking mechanism for bug locations. For each failing test, \bugassist enumerates all diagnoses of a Maximum Satisfiability (MaxSAT) formula corresponding to bug locations. Subsequently, \bugassist ranks diagnoses based on their frequency of appearance in each failing test. 
Other FBFL tools, like \sniper~\cite{jip16-SNIPER}, also enumerate all diagnoses for each failing test. However, the set of \sniper's diagnoses is obtained by taking the Cartesian product of the diagnoses gathered using each failing test.
As a result, while FBFL methods can determine minimal diagnoses per failing test, \bugassist cannot guarantee a minimal diagnosis considering all failing tests, and \sniper may enumerate a significant number of redundant diagnoses that are not minimal~\cite{ijcai19-ignatievMWM}. These limitations may pose challenges for programs with multiple faulty statements, as shown in Example~\ref{eg:motivation}.

\begin{table*}[t]
\begin{minipage}[t!]{0.45\columnwidth}
\centering
\begin{code}
\caption{Faulty program example. Faulty lines: \{5,8,11\}.}
\label{code:motivating_eg}
\begin{minted}[escapeinside=||,tabsize=1,obeytabs,xleftmargin=2pt,linenos]{C}
int main(){
  // finds maximum of 3 numbers
  int f,s,t;
  scanf("%d%d%d",&f,&s,&t);
  if (f < s && f >= t)
    // fix: f >= s
    printf("%d",f);
  if (f > s && s <= t)
    // fix: f < s and s >= t
    printf("%d",s);
  if (f > t && s > t)
    // fix: f < t and s < t
    printf("%d",t);

  return 0;
}
\end{minted}
\end{code}
\end{minipage}
\begin{minipage}[t!]{0.5\columnwidth}
\resizebox{0.8\columnwidth}{!}{%
\begin{tabular}{|l|*{3}{wc{1cm}|}l|wc{1.5cm}|}
\cline{2-4} \cline{6-6}
\multicolumn{1}{c|}{} & \multicolumn{3}{c|}{\textbf{Input}} &  & \textbf{Output} \\ \cline{1-4} \cline{6-6}
{\textbf{$t_0$}} & 1 & 2 & 3 &  & 3 \\ \cline{1-4} \cline{6-6}
{\textbf{$t_1$}} & 6 & 2 & 1 &  & 6 \\ \cline{1-4} \cline{6-6}
{\textbf{$t_2$}} & -1 & 3 & 1 &  & 3 \\ \cline{1-4} \cline{6-6}
\end{tabular}
}
\centering
\caption{Test-suite.}
\label{tab:test-suite}
\hfill
\resizebox{0.85\columnwidth}{!}{%
\begin{tabular}{c|c|c|}
\cline{2-3}
& \textbf{\bugassist}                 & \textbf{\sniper}            \\ \hline
\multicolumn{1}{|c|}{\textbf{\#Diagnoses $t_0$}}  & 8   &  8\\ \hline
\multicolumn{1}{|c|}{\textbf{\#Diagnoses $t_1$}}  &  21  & 21  \\ \hline
\multicolumn{1}{|c|}{\textbf{\#Diagnoses $t_2$}}  & 9  & 9 \\ \hline      
\multicolumn{1}{|c|}{\textbf{\begin{tabular}[c]{@{}c@{}}\#Total\\Unique Diagnoses\end{tabular}}} & 32 & 1297 \\ \hline
\multicolumn{1}{|c|}{\textbf{Final Diagnosis}} & \{4,13\} & \{5,8,11\} \\ \hline
\end{tabular}%
}
\caption{Number of diagnoses (faulty statements) generated by \bugassist~\cite{bugAssist-cav11} and \sniper~\cite{jip16-SNIPER} per test.}
\label{tab:diagnoses}

\end{minipage}
\end{table*}

\begin{example}[Motivation]
\label{eg:motivation}
Consider the program presented in Listing~\ref{code:motivating_eg}, which aims to determine the maximum among three given numbers. However, based on the test suite shown in Table~\ref{tab:test-suite}, the program is faulty, as its output differs from the expected. The set of minimally faulty lines in this program is \{5, 8, 11\}, as all three {\tt if}-conditions are incorrect according to the test suite. Fixing any subset of these lines would be insufficient to repair the program. One possible fix is to replace all these conditions with the suggested fixes in lines \{6, 9, 12\}.

In a typical FBFL approach, the minimal set of statements identified as faulty might include, for example, lines 4 and 5. Removing the \texttt{scanf} statement and an {\tt if}-statement would allow an FBFL tool to assign any value to the input variables in order to always produce the expected output.
However, considering an approach that prioritizes identifying faulty statements within the program's logic before evaluating issues in the input/output statements (such as \texttt{scanf} and \texttt{printf}), one might identify lines \{5, 8, 11\} as the faulty statements.
When applying \bugassist's and \sniper's approach on the program in Listing~\ref{code:motivating_eg} with the described optimization criterion and utilizing the inputs/outputs detailed in Table~\ref{tab:test-suite} as specification, distinct sets of faults are identified for each failing test. Table~\ref{tab:diagnoses} presents the diagnosis (set of faulty lines) produced by each tool, along with the number of diagnoses enumerated for each failing test case and the total number of unique diagnoses after aggregating the diagnoses from all tests, using each tool's respective method.

In the case of \bugassist, diagnoses are prioritized based on their occurrence frequency. Consequently, \bugassist yields 32 unique diagnoses and selects \{4, 13\} since this diagnosis is identified in every failing test. In contrast, \sniper computes the Cartesian product of all diagnoses, resulting in 1297 unique diagnoses.
Note that \bugassist's diagnoses may not adequately identify all faulty program statements. Conversely, \sniper's diagnosis \{5, 8, 11\} is minimal, even though it enumerates an additional 1296 diagnoses.
Hence, existing FBFL methods do not ensure a minimal diagnosis across all failing tests (e.g., \bugassist) or may produce an overwhelming number of redundant sets of diagnoses (e.g., \sniper), especially for programs with multiple faults.
\end{example}

This paper tackles this challenge by formulating the FL problem as a single optimization problem in Section~\ref{sec:mbd-multiple-tests}. We leverage MaxSAT and the theory of \emph{Model-Based Diagnosis (MBD)}, integrating all failing test cases simultaneously. This approach allows us to generate only minimal diagnoses to identify all faulty program components within a C program. 
Furthermore, we have implemented the MBD problem with multiple test cases in \cfaults, a fault localization tool for ANSI-C programs, presented in Section~\ref{sec:cfaults}. \cfaults begins by unrolling and instrumentalizing C programs at the code-level, ensuring independence from the bounded model checker. Next, \cfaults utilizes \cbmc~\cite{tacas04-cbmc-ClarkeKL}, a well-known bounded model checker for C, to generate a trace formula of the program. Finally, \cfaults encodes the problem into MaxSAT to identify the minimal set of diagnoses corresponding to the buggy statements.

Experimental results presented in Section~\ref{sec:results} on two benchmarks of C programs, \tcas~\cite{tcas-dataset-ESE05}~(industrial), and \benchmark~\cite{C-Pack-IPAs_apr24}~(programming exercises), show that \cfaults effectively detects minimal sets of diagnoses. In contrast, \sniper and \bugassist either generate an overwhelming number of redundant diagnoses or fail to produce a minimal set required to fix each program.

To summarize, the contributions of this work are: (1) we tackle the fault localization problem in C programs using a Model-Based Diagnosis (MBD) approach considering multiple failing test cases, and formulating it as a unified optimization problem; 
(2) we implement this MBD approach in a publicly available tool called \cfaults~\cite{CFaults-Zenodo-FM2024}~\footnote{\url{https://github.com/pmorvalho/CFaults}} that unrolls and instrumentalizes C programs at the code level, making it independent of the bounded model checker used; 
(3) \cfaults allows refinement of localized faults to pinpoint the bug's location more precisely;
(4) we evaluate \cfaults on two sets of C programs (\tcas and \benchmark), showing that \cfaults is fast and only produces subset-minimal diagnoses, unlike other state-of-the-art formula-based fault localization~tools.
        
\section{Preliminaries}
\label{sec:prelim}

This section provides definitions and notations that are used throughout the paper.
We start by presenting basic definitions of propositional logic and programs and then address standard \emph{model-based diagnosis (MBD)} definitions.

The \emph{Boolean Satisfiability (SAT)} problem is the decision problem for propositional logic~\cite{biere2009handbook}. 
A propositional formula in Conjunctive Normal Form (CNF) is a conjunction of clauses where each clause is a disjunction of literals.
A literal is a propositional variable $x_i$ or its negation $\neg x_i$.
Given a CNF formula $\phi$, the SAT problem corresponds to deciding if there is an assignment to the variables in $\phi$ such that $\phi$ is satisfied or prove that no such assignment exists.
When applicable, set notation will be used for formulas and clauses. A formula can be represented as a set of clauses (meaning its conjunction) and a clause as a set of literals (meaning its disjunction).

The \emph{Maximum Satisfiability (MaxSAT)} problem is an optimization version of the SAT problem. Given a CNF formula $\phi$, the goal is to find an assignment that maximizes the number of satisfied clauses in $\phi$.
In partial MaxSAT, $\phi$ is split into hard clauses ($\phi_h$) and soft clauses ($\phi_s$). Given a formula $\phi = (\phi_h, \phi_s)$, the goal is to find an assignment that satisfies all hard clauses in $\phi_h$ while minimizing the number of unsatisfied soft clauses in $\phi_s$. Moreover, in the weighted version of the partial MaxSAT problem, each soft clause is assigned a weight, and the goal is to find an assignment that satisfies all hard clauses and minimizes the sum of the weights of the unsatisfied soft clauses. 
Let $\phi = (\phi_h, \phi_s)$ be a partial MaxSAT formula. A Minimal Correction Subset (MCS) $\mu$ of $\phi$ is a subset $\mu \subseteq \phi_s$ where $\phi_h \cup (\phi_s \setminus \mu)$ is satisfiable and, for all $c \in \mu$, $\phi_h \cup (\phi_s \setminus \mu) \cup \lbrace c \rbrace$ is unsatisfiable.
A dual concept of MCSes are \emph{Minimal Unsatisfiable Subsets (MUSes)}~\cite{DBLP:journals/jar/LiffitonS08,ijcai19-ignatievMWM}.

\paragraph{Programs.} A program is considered sequential, comprising standard statements such as assignments, conditionals, loops, and function calls, each adhering to their conventional semantics in C. A program is deemed to contain a bug when an assertion violation occurs during its execution with input $I$. Conversely, if no assertion violation occurs, the program is considered correct for input $I$. In cases where a bug is detected for input $I$, it is possible to define an error trace, representing the sequence of statements executed by program $P$ on input $I$.

A Trace Formula (TF) is a propositional formula that is SAT iff there exists an execution of the program that terminates with a violation of an assert statement while satisfying all assume statements.
For further information on TFs, interested readers are referred to~\cite{tacas04-cbmc-ClarkeKL,DBLP:conf/dac/ClarkeKY03}.

\paragraph{Model-Based Diagnosis (MBD).} The following definitions are commonly used in the \emph{MBD} theory~\cite{reiter87,ijcai19-ignatievMWM,ijcai15-Marques-SilvaJI15}. 
A system description $\mathcal{P}$ is composed of a set of components $\mathcal{C} = \{c_1, \ldots, c_n\}$. Each component in $\mathcal{C}$ can be declared healthy or unhealthy. For each component $c \in \mathcal{C}$, $h(c) = 0$ if $c$ is unhealthy, otherwise, $h(c) = 1$. As in prior works~\cite{ijcai19-ignatievMWM,DBLP:journals/jair/MetodiSKC14}, $\mathcal{P}$ is described by a CNF formula, where $\mathcal{F}_c$ denotes the encoding of component~$c$:

\begin{equation}
    \mathcal{P} \triangleq \bigwedge\nolimits_{c \in \mathcal{C}} { ( \neg h(c) \vee \mathcal{F}_c )}
  \label{eq:system-description}
\end{equation}

Observations represent deviations from the expected system behaviour. An observation, denoted as $o$, is a finite set of first-order sentences~\cite{reiter87,ijcai19-ignatievMWM}, which is assumed to be encodable in CNF as a set of unit clauses. In this work, the failing test cases represent the set of observations.

A system $\mathcal{P}$ is considered faulty if there exists an inconsistency with a given observation $o$ when all components are declared healthy. The problem of model-based diagnosis (MBD) aims to identify a set of components which, if declared unhealthy, restore consistency. This problem is represented by the 3-tuple $\langle \mathcal{P}, \mathcal{C}, o\rangle$, and can be encoded as a CNF formula:

\begin{equation}
    \mathcal{P} \wedge o \wedge \bigwedge\nolimits_{c \in \mathcal{C}} { h(c) } \vDash \bot
  \label{eq:mbd}
\end{equation}

For a given MBD problem $\langle \mathcal{P}, \mathcal{C}, o\rangle$, a set of system components $\Delta \subseteq \mathcal{C}$ is a diagnosis iff:
    \begin{equation}
    \mathcal{P} \wedge o \wedge \bigwedge\nolimits_{c \in \mathcal{C} \setminus \Delta } { h(c) }  \wedge \bigwedge\nolimits_{c \in \Delta} { \neg h(c) } \nvDash \bot
  \label{eq:diagnosis}
  \end{equation}
  
A diagnosis $\Delta$ is minimal iff no subset of $\Delta$, $\Delta' \subsetneq \Delta$, is a diagnosis, and $\Delta$ is of minimal cardinality if there is no other diagnosis $\Delta'' \subseteq \mathcal{C}$ with $|\Delta''| < |\Delta|$.

A diagnosis is redundant if it is not subset-minimal~\cite{ijcai19-ignatievMWM}. 

To encode the Model-Based Diagnosis problem with one observation with partial MaxSAT, the set of clauses that encode $\mathcal{P}$ (\ref{eq:system-description}) represents the set of hard clauses. The soft clauses consists of unit clauses that aim to maximize the set of healthy components, i.e., $\bigwedge_{c \in \mathcal{C}} { h(c) }$~\cite{DBLP:conf/fmcad/SafarpourMVLS07,ijcai15-Marques-SilvaJI15}. This MaxSAT encoding of MBD enables enumerating minimum cardinality diagnoses and subset minimal diagnoses, considering a single observation. 
Furthermore, a minimal diagnosis is a minimal correction subset (MCS) of the MaxSAT formula. Given an inconsistent formula that encodes the MDB problem (\ref{eq:mbd}), a minimal diagnosis $\Delta$ satisfies (\ref{eq:diagnosis}), thereby making $\Delta$ an MCS of the MaxSAT formula. \bugassist~\cite{bugAssist-pldi11}, \sniper~\cite{jip16-SNIPER}, and other model-based diagnosis (MBD) tools for fault localization in circuits~\cite{ijcai15-Marques-SilvaJI15,DBLP:conf/fmcad/SafarpourMVLS07,ijcai19-ignatievMWM} encode the localization problem with partial MaxSAT.

More recently, the MaxSAT encoding for MBD~\cite{ijcai19-ignatievMWM} has been generalized to multiple inconsistent observations.
Let $\mathcal{O} = \{o_1,\ \dots\ o_m\}$ be a set of observations. Each observation is associated with a replica $\mathcal{P}_i$ of the system $\mathcal{P}$. The system remains unchanged given different observations, where the components are replicated for each observation, but the healthy variables are shared. For a given observation $o_i$, a diagnosis is given by the following:
    \begin{equation}
    \mathcal{P}_i \wedge o_i \wedge \bigwedge\nolimits_{c \in \mathcal{C} \setminus \Delta } { h(c) }  \wedge \bigwedge\nolimits_{c \in \Delta} { \neg h(c) } \nvDash \bot
   \label{eq:diagnosis-multiple}
   \end{equation}

The goal is to find a minimal diagnosis $\Delta \subseteq \mathcal{C}$, such that $\Delta$ is a minimal set of components when deactivated the system becomes consistent with all observations $\mathcal{O} = \{o_1,\ \dots\ o_m\}$. Moreover, when considering multiple observations, an aggregated diagnosis is a subset of components that includes one possible diagnosis for each given observation.

\section{Model-Based Diagnosis with Multiple Test Cases}
\label{sec:mbd-multiple-tests}

This paper encodes the fault localization problem as a Model-Based Diagnosis with multiple observations using a single optimization problem. We simultaneously integrate all failing test cases (observations) in a single MaxSAT formula. This approach allows us to generate only minimal diagnoses capable of identifying all faulty components within the system, in our case, a C program. 

Given $m$ observations, $\mathcal{O} = \{o_1, \dots, o_m\}$, a distinct replica of the system, denoted as $\mathcal{P}_i$, is required for each observation $o_i$. The hard clauses, $\phi_h$, in our MaxSAT formulation correspond to each observation's encoding ($o_i$) and $m$ system replicas, one for each observation, $\mathcal{P}_i$. Hence, $\phi_h = \bigwedge_{o_i \in \mathcal{O}} {( \mathcal{P}_i \wedge o_i )}$. Additionally, we aim to maximize the set of healthy components. Therefore, the soft clauses are formulated as: $\phi_s = \bigwedge_{c \in \mathcal{C} } { h(c) }$. Thus, given the MaxSAT solution of $(\phi_h, \phi_s)$, its complement, i.e., the set of unhealthy components  ($h(c)=0$), corresponds to a subset-minimal aggregated diagnosis. This diagnosis is a subset-minimal of components that, when declared unhealthy (deactivated), make the system consistent with all observations, as follows:

\begin{equation}     
     \bigwedge\nolimits_{o_i \in \mathcal{O}} {( \mathcal{P}_i \wedge o_i )} \wedge \bigwedge\nolimits_{c \in \mathcal{C} \setminus \Delta } { h(c) }  \wedge \bigwedge\nolimits_{c \in \Delta} { \neg h(c) } \nvDash \bot
  \label{eq:diagnosis-multiple-single-formula}
\end{equation}

We assume that the system remains unchanged given different observations, where the components are replicated for each observation, but the healthy variables are shared. This is necessary because we analyze all observations jointly, which can affect the component's behaviour.
In our work, the observations consist of a test suite containing failing test cases.

The HSD~\cite{ijcai19-ignatievMWM} algorithm was proposed to localize single faults in circuits given multiple observations. The HSD algorithm is based on hitting set dualization (HSD). For each observation $o_i$, this algorithm computes minimal unsatisfiable subsets (MUSes) of the MaxSAT formula encoded by (\ref{eq:diagnosis-multiple}). Next, the HSD algorithm computes a minimum hitting set $\mathcal{H}$ on the MUSes, and checks if $\mathcal{H}$ makes the system consistent with each observation individually. Hence, to compute all subset-minimal aggregated diagnoses of a faulty system $\mathcal{P}$, the algorithm performs at least $m$ oracle calls for each minimum hitting set computed, where $m$ is the number of observations. Each oracle call uses a different system replica~(\ref{eq:diagnosis-multiple}).

Our approach encodes the problem into a single MaxSAT formula, while HSD~\cite{ijcai19-ignatievMWM} divides the problem into $m$ MaxSAT formulas, one for each observation. Additionally, for each minimal hitting set computed in HSD, $m$ oracle calls are needed to check if a diagnosis is consistent with all observations. However, in our case, we just need to perform a single MaxSAT call that returns a minimal diagnosis, which is, by definition, consistent with all observations since all observations are encoded into the formula. Furthermore, the HSD algorithm was solely evaluated using single faults in circuits given multiple observations, and it was not implemented to work with programs. 
A potential drawback is that our MaxSAT formula grows with the number of observations. This could result in a large formula and affect the performance of the MaxSAT solver. However, this scenario was not observed in our experimental results (see Section~\ref{sec:results}).

\section{\cfaults: MBD with Multiple Observations for C}%
\label{sec:cfaults}

\begin{figure}[t!]
    \centering
    \includegraphics[width=0.8\textwidth]{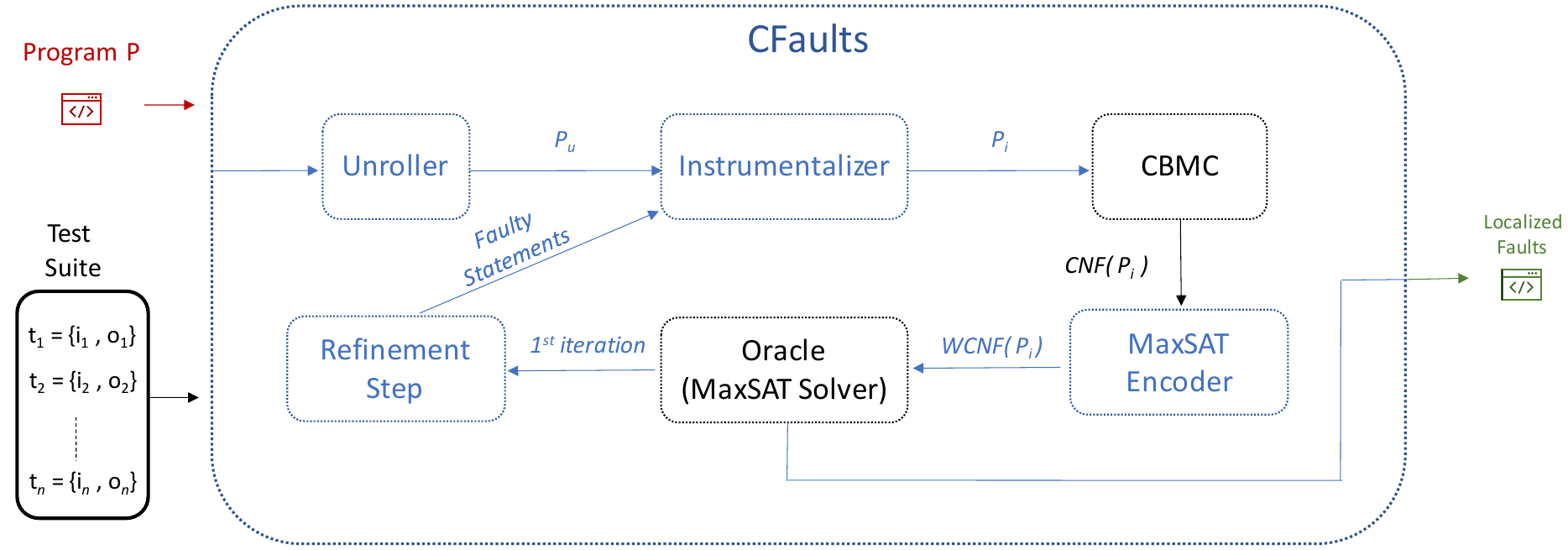}
    \caption{Overview of \cfaults.}
    \label{fig:cfaults-overview}
\end{figure}

\cfaults is a new model-based diagnosis (MBD) tool for fault localization in C programs with multiple test cases. Unlike previous works, \cfaults uses the approach proposed in Section~\ref{sec:mbd-multiple-tests}, and C programs are relaxed at the code level, enabling users to leverage other bounded model checkers effectively.
Figure~\ref{fig:cfaults-overview} provides an overview of \cfaults consisting of six main steps: program unrolling, program instrumentalization, bounded model checking (\cbmc), encoding to MaxSAT, an Oracle (MaxSAT solver), and a refinement step.
Hence, \cfaults formulates the MBD problem with multiple test cases as the 3-tuple $\langle \mathcal{P}, \mathcal{C}, \mathcal{O} \rangle$, where the observations $\mathcal{O}$ consist of failing test cases (inputs and assertions), the components $\mathcal{C}$ represent the set of program statements, and the system description $\mathcal{P}$ is a trace formula of the unrolled and instrumentalized program. The program is instrumented at the code level with relaxation variables corresponding to our \emph{healthy variables}.

\subsubsection{Program unrolling.} 
\cfaults starts the unrolling process by expanding the faulty program using the set of failed tests from the test suite. In this context, an unrolled program signifies the original program expanded $m$ times ($m$ program scopes), where $m$ denotes the number of failed test cases. An unrolled program encodes the execution of all failing tests within the program, along with their corresponding inputs and specifications (assertions).

The unrolling process encompasses three primary steps. Initially, \cfaults generates fresh variables and functions for each of the $m$ program scopes, ensuring each scope possesses unique variables and functions. Subsequently, \cfaults establishes variables representing the inputs and outputs for each program scope corresponding to the failing tests. Input operations, such as \texttt{scanf}, undergo translation into read accesses to arrays corresponding to the inputs, while output operations, such as \texttt{printf}, are replaced by write operations into arrays representing the program's output. Every exit point of the program (e.g., a {\tt return} statement in the {\tt main} function) is replaced with a \texttt{goto} statement directing the program flow to the next failing test's scope. Lastly, at the end of the unrolled program, \cfaults embeds an assertion capturing all the specifications of the failing tests.
Consequently, the unrolled program encapsulates the execution of all failing tests within a single program. 

Listing~\ref{code:prog_unrolled} exhibits a program segment generated through the unrolling process applied to Listing~\ref{code:motivating_eg}. \cfaults establishes global variables to represent the inputs and outputs of each failing test (lines 1--3, Listing~\ref{code:prog_unrolled}). 
For the sake of simplicity, the depicted listing illustrates solely the initial scope corresponding to test 0 from the test suite outlined in Table~\ref{tab:test-suite}. Distinct variables are introduced for each failing test. Furthermore, the \texttt{scanf} function call is substituted with input array operations (lines 8--10), while the \texttt{printf} calls are replaced with \cfaults' print functions, akin to \texttt{sprintf} functions, which direct output to a buffer. Lastly, the unrolled program concludes with an assertion representing the disjunction of the negation of all failing test assertions. For instance, suppose there are $m$ failing tests, where $A_i$ denotes the assertion of test $t_i$. In this scenario, \cfaults injects the following assertion into the program: $\neg A_1 \lor \dots \lor \neg A_m$. 

\begin{figure*}[t]
\begin{code}
\captionof{listing}{The program from Listing~\ref{code:motivating_eg} after being subjected to \cfaults' unrolling process, using the test suite presented in Table~\ref{tab:test-suite}. For simplicity, only the initial scope corresponding to test $t_0$ is displayed. The scopes \texttt{scope\_1} and \texttt{scope\_2} associated with failing tests $t_1$ and $t_2$ are omitted.}
\label{code:prog_unrolled}
\begin{minted}[tabsize=0,obeytabs,xleftmargin=20pt,linenos]{C}
float _input_f0[3] = {1, 2, 3};
char _out_0[2] = "3";
int _ioff_f0 = 0, _ooff_0 = 0;
// ... inputs and outputs for the other tests
int main(){
  scope_0:{
    int f_0, s_0, t_0;
    f_0 = _input_f0[_ioff_f0++];
    s_0 = _input_f0[_ioff_f0++];
    t_0 = _input_f0[_ioff_f0++];
    if ((f_0 < s_0) && (f_0 >= t_0))
        _ooff_0 = printInt(_out_0, _ooff_0, f_0);
    if ((f_0 > s_0) && (s_0 <= t_0))
        _ooff_0 = printInt(_out_0, _ooff_0, s_0);
    if ((f_0 > t_0) && (s_0 > t_0))
        _ooff_0 = printInt(_out_0, _ooff_0, t_0);
    goto scope_1;    
  }
  // ... scope_1 and scope_2
  final_step:
  assert(strcmp(_out_0, "3") != 0 || // other assertions);
}
\end{minted}
\end{code}
\vspace{-.5cm}
\end{figure*}

\subsubsection{Program Intrumentalization.} After integrating all possible executions and assertions from failing tests during the unrolling step, \cfaults proceeds to instrumentalize the unrolled C program by introducing relaxation variables for each program component (statement/instruction). Each relaxation variable activates (or deactivates) the program component being relaxed when assigned to true (or false) respectively. \cfaults ensures that there are no conflicts between the names of the relaxation variables and the names of the program's original variables. For this step, \cfaults needs to receive a maximum number of iterations that the program should be unwound.

The relaxation process introduces relaxation variables that deactivate or activate program components. This process involves four distinct relaxation rules for: (1) conditions of {\tt if}-statements, (2) expression lists (e.g., an expression list executed at the beginning of a for-loop), (3) loop conditions, and (4) other program statements.

\begin{example}
\label{eg:code-relaxation}
Listings~\ref{code:prog_statements} shows a code snippet that sums all the numbers between $1$ and \texttt{n}. Listings~\ref{code:statements_relaxed} depicts the same program statements after undergoing relaxation by \cfaults. For the sake of simplicity, all relaxation variables' and offsets' names were simplified.
\end{example}

\begin{table*}[t]
\begin{minipage}[t!]{0.45\columnwidth}
\begin{code}
\vspace{-.88in}
\captionof{listing}{Program statements.}
\label{code:prog_statements}
\begin{minted}[escapeinside=||,tabsize=0,obeytabs,xleftmargin=20pt,linenos]{C}
int i;
int n;
int s;

s = 0;
n = _input_f0[_ioff_f0++];

if (n == 0)
    return 0;

for (i=1; i < n; i++){
    s = s + i;
}
\end{minted}
\end{code}
\end{minipage}
\begin{minipage}[t!]{0.54\columnwidth}
\begin{code}
\captionof{listing}{Program statements relaxed.}
\label{code:statements_relaxed}
\begin{minted}[escapeinside=çç,tabsize=1,obeytabs,xleftmargin=20pt,linenos]{C}
//main scope
bool ç\textcolor{blue}{\textit{\_rv1, \_rv2, \_rv3, \_rv5}}ç;
bool ç\textcolor{blue}{\textit{\_rv6[UNWIND],..., \_rv8[UNWIND]}}ç;
int ç\textcolor{blue}{\textit{\_los}}ç; // loop1 offset

//test scope 
bool ç\textcolor{blue}{\textit{\_ev4}}ç;
int i,n,s;
ç\textcolor{blue}{\textit{\_los=1}}ç;

if (ç\textcolor{blue}{\textit{\_rv1}}ç) s = 0;
if (ç\textcolor{blue}{\textit{\_rv2}}ç) n = _input_f0[_ioff_f0++];

if ( ç\textcolor{blue}{\textit{\_rv3}}ç ? (n == 0) : ç\textcolor{blue}{\textit{\_ev4}}ç)
    return 0;

for (ç\textcolor{blue}{\textit{\_rv5}}ç ? (i = 1) : ç\textcolor{blue}{\textit{1}}ç; 
     ç\textcolor{blue}{\textit{!\_rv6[\_los]}}ç || (i<n); 
     ç\textcolor{blue}{\textit{\_rv8[\_los]}}ç ? i++ : ç\textcolor{blue}{\textit{1, \_los++}}ç){
    if (ç\textcolor{blue}{\textit{\_rv7[\_los]}}ç) s = s + i;       
}
\end{minted}
\end{code}
\end{minipage}
\vspace{-0.15in}
\end{table*}

In more detail, the rule for relaxing a general program statement is to envelop the statement with an {\tt if}-statement, whose condition is a relaxation variable. For example, consider lines 5 and 6 in the program on Listings~\ref{code:prog_statements}. These lines are relaxed by \cfaults using relaxation variables \texttt{\_rv1} and \texttt{\_rv2} respectively, appearing as lines 11 and 12 on Listings~\ref{code:statements_relaxed}.

Furthermore, when relaxing {\tt if}-statements, the statements inside the {\tt then} and {\tt else} blocks adhere to the previously explained relaxation rule. However, the conditions of {\tt if}-statements are relaxed using a ternary operator, as shown in line 14 of Listings~\ref{code:statements_relaxed}. Note that if the relaxation variable is assigned true, then the original {\tt if} condition is executed. Otherwise, a different relaxation variable (e.g., \texttt{\_ev4} in Listings~\ref{code:statements_relaxed}) determines whether the program execution enters the {\tt then}-block or the {\tt else}-block (if one exists).
These relaxation variables (\emph{{\tt else}'s relaxation variables}) are local to each failing test scope and enable different tests to determine whether to enter the {\tt then} or {\tt else}-block.

When handling expression lists, \cfaults adopts a comparable strategy to that of generic program statements, enclosing each expression within a ternary operator instead of an {\tt if}-statement. If the program component is deactivated, the expression is replaced by \texttt{1}. For example, the initialization of variable \texttt{i} in line 11 of Listings~\ref{code:prog_statements} is relaxed into the ternary operation in line 17 of Listings~\ref{code:statements_relaxed}.

Lastly, all relaxation variables inside a loop are Boolean vectors to relax statements within a loop. Each entry of these vectors relaxes the loop's statements for a given iteration. The maximum number of iterations of the loops is defined by the \cfaults user.
\cfaults follows a similar approach for inner loops, creating arrays of arrays. Thus, for simple program statements within a loop, \cfaults encapsulates them with {\tt if}-statements, with the relaxation variables indexed to the iteration number. Line 20 of Listings~\ref{code:statements_relaxed} illustrates a relaxed statement inside a loop. The loop's condition is relaxed by implication of the relaxation variable, as demonstrated in line 18 of Listings~\ref{code:statements_relaxed}. Furthermore, each loop has its own offsets to index relaxation variables. These offsets are initialized just before the loop and incremented at the end of each iteration (e.g., line 19 in Listing~\ref{code:statements_relaxed}).

When handling auxiliary functions, \cfaults declares the relaxation variables needed in the main scope of the program and passes these variables as parameters. Hence, \cfaults ensures that the same variables are used throughout the auxiliary functions' calls.

Listing~\ref{code:prog_inst} depicts the program resulting from the instrumentalization process of Listing~\ref{code:prog_unrolled} performed by \cfaults. 
The same program components (statements/instructions) across different failing test scopes are assigned the same relaxation variable declared in the main scope. Consequently, if a relaxation variable is set to 0, the corresponding program component is deactivated across all test executions.  Additionally, the relaxation variables are left uninitialized, allowing \cfaults to determine the minimal number of faulty components requiring deactivation. Note that relaxation variables are not declared as global variables but as local variables within the {\tt main} scope. This is to prevent the C compiler from automatically initializing all these variables to 0.

\begin{figure*}[t]
\begin{code}
\captionof{listing}{Instrumentalized program.}
\label{code:prog_inst}
\begin{minted}[escapeinside=çç,tabsize=1,obeytabs,xleftmargin=20pt,linenos]{C}
//global vars
int main(){
  bool ç\textcolor{blue}{\textit{\_rv1}}ç, ç\textcolor{blue}{\textit{\_rv2}}ç, ..., ç\textcolor{blue}{\textit{\_rv12}}ç;
  scope_0:{
    bool ç\textcolor{blue}{\textit{\_ev5}}ç, ç\textcolor{blue}{\textit{\_ev8}}ç, ç\textcolor{blue}{\textit{\_ev11}}ç;
    int f_0, s_0, t_0;
    if (ç\textcolor{blue}{\textit{\_rv1}}ç) f_0 = _input_f0[_ioff_f0++];
    if (ç\textcolor{blue}{\textit{\_rv2}}ç) s_0 = _input_f0[_ioff_f0++];
    if (ç\textcolor{blue}{\textit{\_rv3}}ç) t_0 = _input_f0[_ioff_f0++];
    if (ç\textcolor{blue}{\textit{\_rv4}}ç ? ((f_0 < s_0) && (f_0 >= t_0)) : ç\textcolor{blue}{\textit{\_ev5}}ç ){
        if (ç\textcolor{blue}{\textit{\_rv6}}ç) _ooff_0 = printInt(_out_0, _ooff_0, f_0);
    }
    if (ç\textcolor{blue}{\textit{\_rv7}}ç ? ((f_0 > s_0) && (s_0 <= t_0)) : ç\textcolor{blue}{\textit{\_ev8}}ç ){
        if (ç\textcolor{blue}{\textit{\_rv9}}ç) _ooff_0 = printInt(_out_0, _ooff_0, s_0);
    }
    if (ç\textcolor{blue}{\textit{\_rv10}}ç? ((f_0 > t_0) && (s_0 > t_0)) : ç\textcolor{blue}{\textit{\_ev11}}ç ){
        if (ç\textcolor{blue}{\textit{\_rv12}}ç) _ooff_0 = printInt(_out_0, _ooff_0, t_0);
    }
    goto scope_1;    
  }
  // scope_1 and scope_2
  final_step:
  assert(strcmp(_out_0, "3") != 0 || ... // other assertions);
}
\end{minted}
\end{code}
\end{figure*}

\subsubsection{\cbmc.}
After unrolling and instrumentalizing the C program, \cfaults invokes \cbmc, a bounded model checker for C~\cite{tacas04-cbmc-ClarkeKL}. \cbmc initially transforms the unrolled and relaxed program into \emph{Static Single Assignment (SSA)} form, an intermediate representation ensuring that variables are assigned values only once and are defined before use~\cite{ssa-tpls91}. SSA achieves this by converting existing variables into multiple versions, each uniquely representing an assignment. Next, \cbmc translates the SSA representation into a CNF formula, which represents the trace formula of the program.
During the CNF formula generation, \cbmc negates the program's assertion ($\neg (\neg A_1 \lor \dots \lor \neg A_m)$) to compute a counter-example. Moreover, the CNF formula, $\phi$, encodes each failing test's input ($I_i$), assertion ($A_i$), and all execution paths of the unrolled and relaxed incorrect program encoded by the trace formula ($P$), i.e., $\phi = (I_1\ \land\ \dots\ \land\ I_m)\ \land\ P\ \land\ (A_1 \land \dots \land A_m)$. Thus, if $\phi$ is $SAT$, an assignment exists that activates or deactivates each relaxation variable and makes all failing test assertions true. Hence, each satisfiable assignment is a diagnosis of the C program, considering all failing tests. 

\subsubsection{MaxSAT Encoder.} 
Let $\phi$ denote the CNF formula generated by \cbmc in the previous step. Next, \cfaults generates a weighted partial MaxSAT formula $(\mathcal{H}, \mathcal{S})$ to maximize the satisfaction of relaxation variables in the program, aiming to minimize the necessary code alterations. The set of hard clauses is defined by \cbmc's CNF formula (i.e., $\mathcal{H} = \phi$), while the soft clauses consist of unit clauses representing relaxation variables used to instrument the C program, expressed as $\mathcal{S} = \bigwedge_{c \in \mathcal{C}} { ({rv}_c) }$. Additionally, we assign a hierarchical weight to each relaxation variable based on the height of its sub-AST (Abstract Syntax Tree). For instance, in the case of an {\tt if}-statement without an {\tt else}-block, the relaxation variable for its condition will be assigned a weight equal to the sum of the weights of the relaxation variables within the {\tt then}-block. Furthermore, to prioritize the identification of faulty statements within the program's logic over evaluating issues in the input/output, these statements (such as \texttt{scanf} and \texttt{printf}) are assigned a significantly higher cost compared to other program statements.
Moreover, due to the use of hierarchical weights in the relaxation variables, \cfaults enumerates all MaxSAT solutions to identify all subset-minimal diagnoses since there can be more than one MaxSAT solution (with the same cost) that differ in the number of relaxed program statements.

\subsubsection{Oracle.} \cfaults invokes a MaxSAT solver to determine the program's minimal set of faulty statements, aligning with the principles of Model-Based Diagnosis (MBD) theory. By consolidating all failing tests into a unified, unrolled, and instrumentalized program, the MaxSAT solution identifies the minimum subset of statements requiring removal to fulfil the assertions of all failing tests.

\subsubsection{Refinement.} The standard Model-Based Diagnosis (MBD) theory focuses on faulty components (program statements) whose removal can rectify the system (program's assertions). However, addressing program faults in software may necessitate introducing, relocating, or replacing statements. Hence, \cfaults incorporates a refinement step that introduces nondeterminism into the program, enabling the Oracle to simulate actions such as introducing, reallocating or replacing existing program statements. During the first iteration of \cfaults, 
the refinement step is invoked to introduce non-determinism, with the aim of minimizing the number of faulty statements. 
This step can improve fault localization by conducting a more detailed analysis of previously identified faulty statements.
For example, in the scenario outlined in Example~\ref{eg:motivation}, refining line 5 into
\begin{code}
\begin{minted}[escapeinside=çç,tabsize=1,obeytabs,xleftmargin=0pt,fontsize=\small]{C}
if ((ç\textcolor{blue}{\textit{\_rv1?}}ç (f < s) ç\textcolor{blue}{\textit{: nondet\_bool()}}ç) && (ç\textcolor{blue}{\textit{\_rv2?}}ç (f >= t) ç\textcolor{blue}{\textit{: nondet\_bool()}}ç))
\end{minted}
\end{code}
enables \cfaults to determine that only the left part of the binary operation (\texttt{f < s}) is faulty, while the right part remains unaffected. This fine-grained approach allows for more precise detection of program faults.
When the refinement step is triggered, \cfaults instrumentalizes the program again, introducing nondeterminism exclusively to the statements previously identified as faulty during the initial Oracle call. Through this process, \cfaults aims to reduce the set of faulty program components by executing them or assigning them to nondeterministic functions. All remaining program components are executed, meaning their relaxation variables are activated during this step.
 
\section{Experimental Results}
\label{sec:results}

All of the experiments were conducted on an Intel(R) Xeon(R) Silver computer with
4210R CPUs @ 2.40GHz running Linux Debian 10.2, using a memory limit of 32 GB and a timeout of 3600s, for each program.
\cfaults has been evaluated using two distinct benchmarks of C programs: \tcas~\cite{tcas-dataset-ESE05} and \benchmark~\cite{C-Pack-IPAs}. \tcas stands out as a well-known program benchmark extensively utilized in the fault localization literature~\cite{bugAssist-pldi11,jip16-SNIPER}. This benchmark comprises a C program from Siemens and 41 versions with intentionally introduced faults, with known positions and types of these faults. Conversely, \benchmark is a set of student programs collected during an introductory programming course. For this evaluation, we used the first lab class of \benchmark, which consists of ten programming assignments, comprising 486 faulty programs and 799 correct implementations. \benchmark has proven successful in evaluating various works across program analysis~\cite{ecai23-GNNs-4-var-mapping}, program transformation~\cite{fse22-MultIPAS}, and clustering~\cite{InvAASTCluster-corr22}.

\cfaults uses \texttt{pycparser}~\cite{pycparser}
for unrolling and instrumentalizing C programs. Additionally, \cbmc version 5.11 is used to encode C programs into CNF formulas. Furthermore, since the source code of \bugassist and \sniper is either unavailable or no longer maintained (resulting in compilation and linking issues), prototypes of their algorithms were implemented. It is worth noting that the original version of \sniper could only analyze programs that utilized a subset of ANSI-C, lacked support for loops and recursion, and could only partially handle global variables, arrays, and pointers. In this work, both \sniper and \bugassist handle ANSI-C programs, as their algorithms are built on top of \cfaults's unroller and instrumentalizer modules.
For the MaxSAT oracle, RC2Stratified~\cite{imms19-RC2} from the \texttt{PySAT} toolkit~\cite{imms18-PySAT} (v. 0.1.7.dev19) was used.

Furthermore, all three FBFL algorithms evaluated (\cfaults, \bugassist, and \sniper) consistently generate diagnoses that are consistent with (\ref{eq:diagnosis-multiple-single-formula}), indicating that all proposed diagnoses undergo validation by \cbmc once the algorithm provides a diagnosis. However, this validation primarily serves to verify diagnoses generated by \bugassist, as it has the capability to produce diagnoses that may not align with all failing test cases. In contrast, \cfaults' MaxSAT solution, by definition, aligns with all observations, and \sniper's aggregation method (Cartesian product) produces only valid diagnoses, although they may not always be subset-minimal. When considering \bugassist, we iterate through all computed diagnoses based on \bugassist's voting score, until we identify one diagnosis that is consistent with all observations, i.e., conforms to (\ref{eq:diagnosis-multiple-single-formula}).

\begin{table*}[t!]
\begin{minipage}[t!]{0.49\columnwidth}
\resizebox{\columnwidth}{!}{%
\begin{tabular}{ccccccc}
\multicolumn{7}{c}{Benchmark: \textbf{TCAS}} \\ \hline
{} && \begin{tabular}[c]{@{}c@{}}\textbf{Valid}\\\textbf{Diagnosis}\end{tabular} &&  \textbf{Memouts} && \textbf{Timeouts}\\
\hline
\textbf{BugAssist} & & 41 (100.0\%) & & 0 (0.0\%) & & 0 (0.0\%)\\
\textbf{SNIPER} & & 7 (17.07\%) & & 34 (82.93\%) & & 0 (0.0\%)\\
\textbf{CFaults} & & 41 (100.0\%) & & 0 (0.0\%) & & 0 (0.0\%)\\
\textbf{CFaults-Refined} & & 41 (100.0\%) & & 0 (0.0\%) & & 0 (0.0\%)\\
\bottomrule
\end{tabular}
}
\end{minipage}
\hfill
\begin{minipage}[t!]{0.49\columnwidth}
\resizebox{\columnwidth}{!}{%
\begin{tabular}{ccccccc}

\multicolumn{7}{c}{Benchmark: \textbf{C-Pack-IPAs}} \\ \hline
{} && \begin{tabular}[c]{@{}c@{}}\textbf{Valid}\\\textbf{Diagnosis}\end{tabular} &&  \textbf{Memouts} && \textbf{Timeouts}\\
\hline
\textbf{BugAssist} & & 454 (93.42\%) & & 0 (0.0\%) & & 32 (6.58\%)\\
\textbf{SNIPER} & & 446 (91.77\%) & & 4 (0.82\%) & & 36 (7.41\%)\\
\textbf{CFaults} & & 483 (99.38\%) & & 1 (0.21\%) & & 2 (0.41\%)\\
\textbf{CFaults-Refined} & & 482 (99.18\%) & & 1 (0.21\%) & & 3 (0.62\%)\\
\bottomrule
\end{tabular}
}
\end{minipage}
\caption{\bugassist, \sniper and \cfaults fault localization results.}
\label{tab:fault_loc_results}
\vspace{-0.15in}
\end{table*}

\begin{figure*}[t!]
    \begin{subfigure}[t!]{0.48\textwidth}
    \centering
    \includegraphics[width=0.75\textwidth]{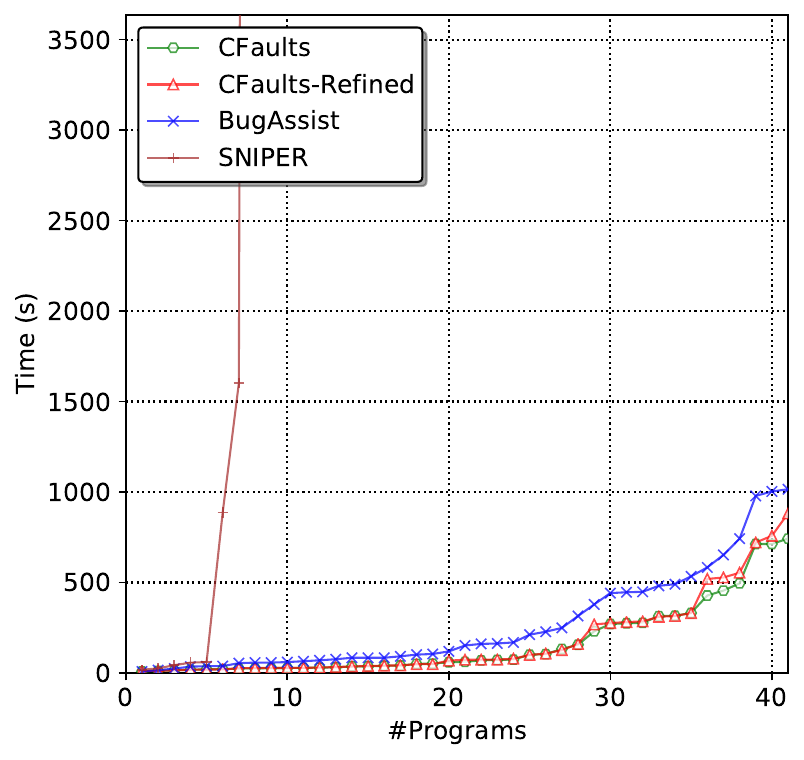} 
     \caption{Time Performance on \tcas.}
     \label{fig:tcas-cpu_time}
     \end{subfigure}
     \hspace{0.02\textwidth}
    \begin{subfigure}[t!]{0.48\textwidth}
    \centering
    \includegraphics[width=0.75\textwidth]{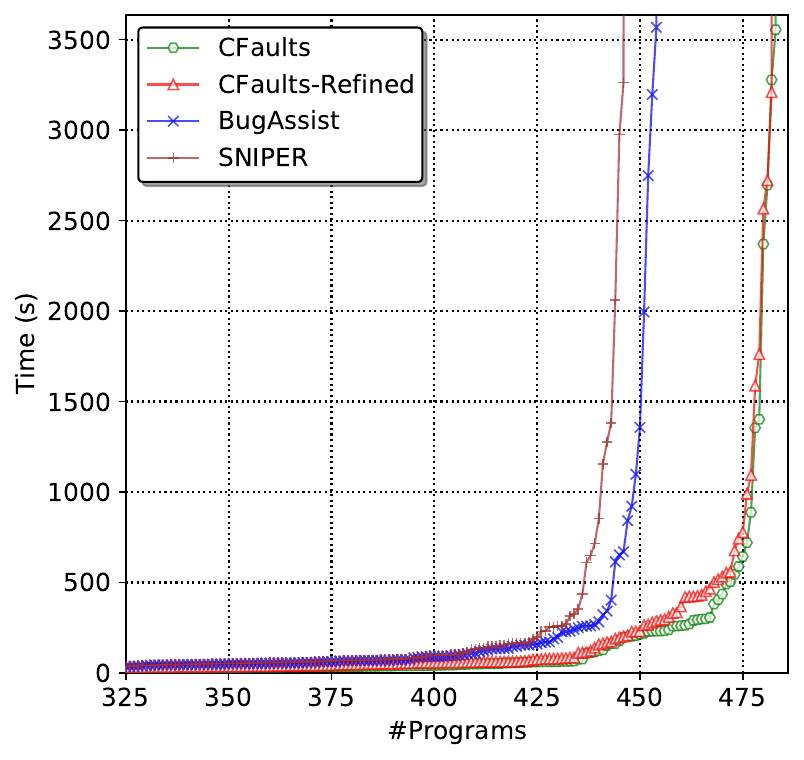}
     \caption{Time Performance on \benchmark.}
     \label{fig:cpackipas-cpu_time}
     \end{subfigure}

    \begin{subfigure}[t!]{0.32\textwidth}    
    \centering
    \includegraphics[width=\textwidth]{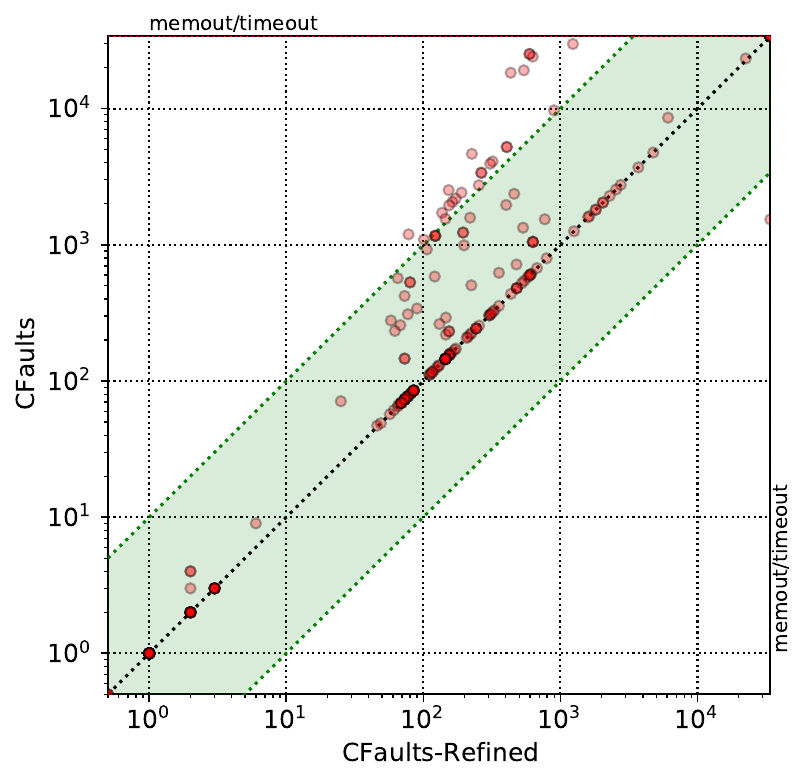}
    \caption{Costs of refined
    diagnoses 
    on \benchmark.}
    \label{fig:scatter-refinement}
     \end{subfigure}
    \begin{subfigure}[t!]{0.32\textwidth}    
    \includegraphics[width=\textwidth]{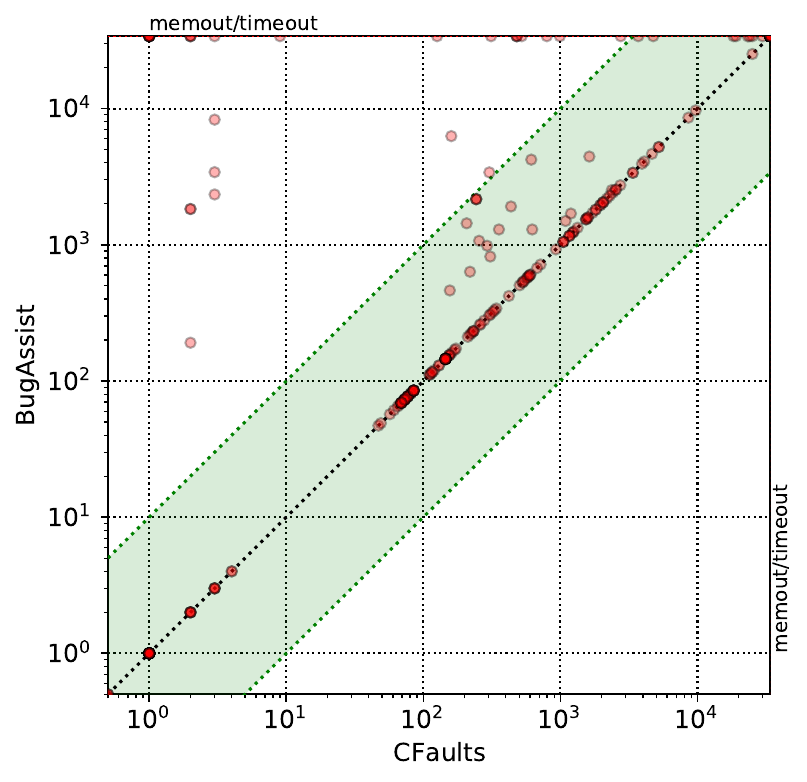}
    \caption{Costs of 
    diagnoses on \benchmark.}
    \label{fig:scatter-costs}
     \end{subfigure}
    \begin{subfigure}[t!]{0.32\textwidth}    
    \includegraphics[width=\textwidth]{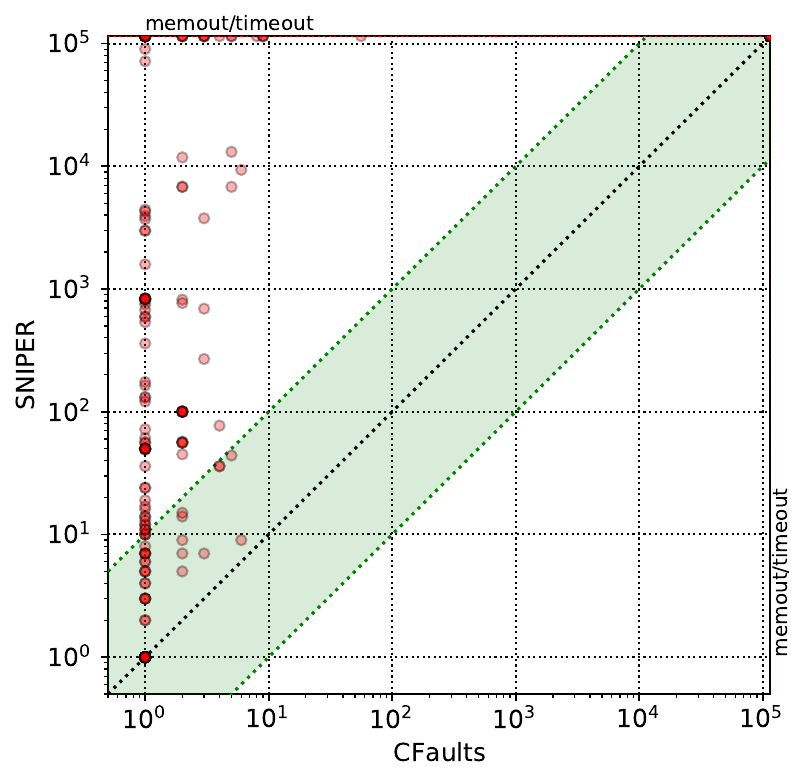}
    \caption{\#Diagnoses generated 
    on \benchmark.}
    \label{fig:scatter-num-diagnoses}
     \end{subfigure}
    \caption{Comparison between \bugassist's, \sniper's and \cfaults' diagnoses.}
    \label{fig:plots}
\end{figure*}

Table~\ref{tab:fault_loc_results} provides an overview of the results obtained using \sniper, \bugassist, and \cfaults on the two benchmarks of C programs. The \tcas program comprises approximately 180 lines of code and has a maximum of 131 failing tests for each program. This leads \sniper to reach the memory limit of 32GB for almost 83\% of the programs when aggregating the sets of MCSes computed for each failing test. Additionally, a higher rate of timeouts is observed for \sniper and \bugassist than for \cfaults.
Figures~\ref{fig:tcas-cpu_time}~and~\ref{fig:cpackipas-cpu_time} depict cactus plots that present the CPU time spent on fault localization in each program (y-axis) versus the number of programs with all faults successfully localized (x-axis) using \bugassist, \sniper, and \cfaults (with and without refinement) on \tcas and \benchmark, respectively. Notably, \cfaults generally exhibits faster performance compared to \bugassist and \sniper across both benchmarks. 
In Figure~\ref{fig:tcas-cpu_time}, \sniper's performance is due to its memout rate on \tcas.

In \tcas, \cfaults, whether invoking the refinement step or not, identifies faults in the entire dataset. However, in \benchmark, \cfaults localizes faults in one additional program when the refinement step is not called. Even if the refinement step reaches the time limit, \cfaults still possesses a subset-minimal diagnosis from the preceding step that has not undergone refinement.
The refinement step slightly slows down \cfaults, as shown in Figures~\ref{fig:tcas-cpu_time} and \ref{fig:cpackipas-cpu_time}.
Nonetheless, Figure~\ref{fig:scatter-refinement} illustrates a scatter plot comparing the optimum costs (MaxSAT solution's cost) achieved by \cfaults with and without calling the refinement step on \benchmark. Each point on this plot represents a faulty program, where the x-value (resp. y-value) represents the optimum cost of \cfaults' with refinement (resp. without refinement) diagnosis. If a point lies above the diagonal, it indicates that a non-refined diagnosis has a higher cost than a refined diagnosis for the same program. Therefore, while the refinement step may marginally slow down \cfaults, it enables \cfaults to identify smaller diagnoses at a reduced cost in approximately 16\% of \benchmark's programs. Moreover, this observation was not noted in the \tcas dataset, as each program contains a maximum of two faults, and the refinement step did not yield improved outcomes in this particular dataset.

Additionally, Figure~\ref{fig:scatter-costs} illustrates a scatter plot comparing the diagnoses' costs achieved by \cfaults (x-axis) against \bugassist (y-axis) on \benchmark. 
\bugassist fails to provide an optimal diagnosis in almost 6\% of cases. In the \tcas benchmark, although \bugassist manages to localize faults in all programs, it yields a non-optimal diagnosis in 10\% of the programs. Furthermore, Figure~\ref{fig:scatter-num-diagnoses} depicts a scatter plot comparing the number of diagnoses generated by \cfaults (x-axis) against \sniper (y-axis). While \cfaults needs to enumerate all MaxSAT solutions due to the weighted MaxSAT formula, it is evident that \sniper generates significantly more diagnoses than \cfaults. This discrepancy suggests that \sniper overlooks the possibility of redundant diagnoses being computed. The number of such redundant diagnoses is much larger than the subset-minimal diagnoses generated by \cfaults. Figure~\ref{fig:scatter-num-diagnoses} illustrates that in some instances, \sniper may enumerate up to 100K diagnoses, whereas \cfaults generates less than 10.

As a validation step for our implementation, we analyzed all three fault localization methods on the collection of 799 correct programs in \benchmark. This was done to ensure that all methods yielded zero faults for all correct implementations of each programming exercise.
Moreover, we conducted a comparison between \cfaults and the HSD algorithm~\cite{ijcai19-ignatievMWM} (see Section~\ref{sec:mbd-multiple-tests}) on the \textsc{ISCAS85} dataset~\cite{iscas85}, which is a widely studied collection of single-fault circuits. It is worth noting that HSD's implementation currently only supports fault localization in circuits. We encountered no performance issues during this comparison, and both approaches successfully localized all faults within each circuit.

\section{Related Work}

Fault localization (FL) techniques typically fall into two main families:
\emph{spectrum-based (SBFL)} and \emph{formula-based (FBFL)}. SBFL methods~\cite{DBLP:conf/kbse/AbreuZG09,DBLP:journals/jss/WongDC10,DBLP:journals/tosem/NaishLR11,DBLP:journals/tr/WongDGL14,DBLP:journals/tse/WongGLAW16,DBLP:journals/jss/AbreuZGG09} estimate the likelihood of a statement being faulty based on test coverage information from both passing and failing test executions. While SBFL techniques are generally fast, they may lack precision, as not all identified statements are likely to be the cause of failures~\cite{liu2019you,DBLP:conf/cav/RothenbergG20}.
In contrast, FBFL approaches~\cite{bugAssist-cav11,bugAssist-pldi11,jip16-SNIPER,lambda2,DBLP:conf/icfem/LamraouiN14,DBLP:journals/entcs/GriesmayerSB07,DBLP:journals/jlp/WotawaNM12,DBLP:conf/popl/XieA05,DBLP:conf/fmcad/KonighoferB11} are considered exact. FBFL methods encode the fault localization problem into several optimization problems aimed at identifying the minimum number of faulty statements within a program. Typically, these methods perform a MaxSAT call for each failing test, allowing them to individually identify a minimal set of faults for each failing test case rather than simultaneously addressing all failing test cases.
\emph{Program slicing}~\cite{DBLP:journals/ese/SoremekunKBZ21,DBLP:conf/cav/RothenbergG20,zeller1999yesterday}  has also emerged as a technique for localizing faults within programs. A more syntactic FBFL approach~\cite{DBLP:conf/cav/RothenbergG20} is to use program slicing to enumerate all minimal sets of repairs for a given faulty program. Another method for identifying the causes of faulty program behaviour involves analyzing the variances between various versions of the software~\cite{zeller1999yesterday}. 
\emph{Refinement} has a long-standing tradition in verification; 
particularly for refining abstractions of reachable states~\cite{DBLP:conf/tacas/ClarkeKSY05,DBLP:journals/fmsd/ClarkeKSY04,DBLP:books/daglib/0007403-2}. In that sense, our form of refinement is different because it enables us to more precisely pinpoint faults of the user, at the sub-expression level.

\section{Conclusion}

This paper introduces a novel formula-based fault localization technique for C programs capable of addressing any number of faults. Leveraging Model-Based Diagnosis (MBD) with multiple observations, \cfaults consolidates all failing test cases into a unified MaxSAT formula, ensuring consistency in the fault localization process.
Experimental evaluations on \tcas and \benchmark, show that \cfaults is faster than other FBFL approaches like \bugassist and \sniper. Furthermore, \cfaults only generates minimal diagnoses of faulty statements, while other methods tend to produce redundant diagnoses.

\section*{Acknowledgements}
This work was partially supported by Portuguese national funds through FCT, under projects UIDB/50021/2020 (DOI: 10.54499/\-UIDB/\-50021/\-2020), PTDC/\-CCI-COM/\-2156/2021 (DOI: 10.54499/\-PTDC/\-CCI-COM/\-2156/\-2021) and 2022.\-03537.PTDC (DOI: 10.54499/\-2022.03537.PTDC) and grant SFRH/\-BD/\-07724/\-2020 (DOI: 10.54499/\-2020.07724.BD). PO acknowledges travel support from the European Union’s Horizon 2020 research and innovation programme under ELISE Grant Agreement No 951847.
This work was also supported by the MEYS within the program ERC CZ under the project POSTMAN no.~LL1902 and co-funded by the European Union under the project \emph{ROBOPROX} (reg.~no.~CZ\-.02.01.01/00/\-22\_008/0004590). This article is part of the RICAIP project that has received funding from the EU’s Horizon 2020 research and innovation program under grant agreement No 857306.

%
%
%
\bibliographystyle{splncs04}
\bibliography{mybibliography}

\begin{thebibliography}{10}
\providecommand{\url}[1]{\texttt{#1}}
\providecommand{\urlprefix}{URL }
\providecommand{\doi}[1]{https://doi.org/#1}

\bibitem{DBLP:conf/kbse/AbreuZG09}
Abreu, R., Zoeteweij, P., van Gemund, A.J.C.: Spectrum-based multiple fault localization. In: {ASE} 2009, 24th {IEEE/ACM} International Conference on Automated Software Engineering, Auckland, New Zealand, November 16-20, 2009. pp. 88--99. {IEEE} Computer Society (2009). \doi{10.1109/ASE.2009.25}, \url{https://doi.org/10.1109/ASE.2009.25}

\bibitem{DBLP:journals/jss/AbreuZGG09}
Abreu, R., Zoeteweij, P., Golsteijn, R., van Gemund, A.J.C.: A practical evaluation of spectrum-based fault localization. J. Syst. Softw.  \textbf{82}(11),  1780--1792 (2009). \doi{10.1016/J.JSS.2009.06.035}, \url{https://doi.org/10.1016/j.jss.2009.06.035}

\bibitem{biere2009handbook}
Biere, A., Heule, M., van Maaren, H., Walsh, T. (eds.): Handbook of Satisfiability, Frontiers in Artificial Intelligence and Applications, vol.~185. {IOS} Press (2009)

\bibitem{DBLP:books/daglib/0007403-2}
Clarke, E.M., Grumberg, O., Kroening, D., Peled, D.A., Veith, H.: Model checking, 2nd Edition. {MIT} Press (2018), \url{https://mitpress.mit.edu/books/model-checking-second-edition}

\bibitem{tacas04-cbmc-ClarkeKL}
Clarke, E.M., Kroening, D., Lerda, F.: A tool for checking {ANSI-C} programs. In: Jensen, K., Podelski, A. (eds.) Tools and Algorithms for the Construction and Analysis of Systems, 10th International Conference, {TACAS} 2004, Held as Part of the Joint European Conferences on Theory and Practice of Software, {ETAPS} 2004, Barcelona, Spain, March 29 - April 2, 2004, Proceedings. Lecture Notes in Computer Science, vol.~2988, pp. 168--176. Springer (2004). \doi{10.1007/978-3-540-24730-2\_15}, \url{https://doi.org/10.1007/978-3-540-24730-2\_15}

\bibitem{DBLP:journals/fmsd/ClarkeKSY04}
Clarke, E.M., Kroening, D., Sharygina, N., Yorav, K.: Predicate abstraction of {ANSI-C} programs using {SAT}. Formal Methods Syst. Des.  \textbf{25}(2-3),  105--127 (2004). \doi{10.1023/B:FORM.0000040025.89719.F3}, \url{https://doi.org/10.1023/B:FORM.0000040025.89719.f3}

\bibitem{DBLP:conf/tacas/ClarkeKSY05}
Clarke, E.M., Kroening, D., Sharygina, N., Yorav, K.: {SATABS:} sat-based predicate abstraction for {ANSI-C}. In: Halbwachs, N., Zuck, L.D. (eds.) Tools and Algorithms for the Construction and Analysis of Systems, 11th International Conference, {TACAS} 2005, Held as Part of the Joint European Conferences on Theory and Practice of Software, {ETAPS} 2005, Edinburgh, UK, April 4-8, 2005, Proceedings. Lecture Notes in Computer Science, vol.~3440, pp. 570--574. Springer (2005). \doi{10.1007/978-3-540-31980-1\_40}, \url{https://doi.org/10.1007/978-3-540-31980-1\_40}

\bibitem{DBLP:conf/dac/ClarkeKY03}
Clarke, E.M., Kroening, D., Yorav, K.: Behavioral consistency of {C} and verilog programs using bounded model checking. In: Proceedings of the 40th Design Automation Conference, {DAC} 2003, Anaheim, CA, USA, June 2-6, 2003. pp. 368--371. {ACM} (2003). \doi{10.1145/775832.775928}, \url{https://doi.org/10.1145/775832.775928}

\bibitem{ssa-tpls91}
Cytron, R., Ferrante, J., Rosen, B.K., Wegman, M.N., Zadeck, F.K.: Efficiently computing static single assignment form and the control dependence graph. {ACM} Trans. Program. Lang. Syst.  \textbf{13}(4),  451--490 (1991). \doi{10.1145/115372.115320}, \url{https://doi.org/10.1145/115372.115320}

\bibitem{tcas-dataset-ESE05}
Do, H., Elbaum, S.G., Rothermel, G.: Supporting controlled experimentation with testing techniques: An infrastructure and its potential impact. Empir. Softw. Eng.  \textbf{10}(4),  405--435 (2005). \doi{10.1007/S10664-005-3861-2}

\bibitem{lambda2}
Feser, J.K., Chaudhuri, S., Dillig, I.: Synthesizing data structure transformations from input-output examples. In: Proceedings of the 36th {ACM} {SIGPLAN} Conference on Programming Language Design and Implementation, Portland, OR, USA, June 15-17, 2015. pp. 229--239 (2015)

\bibitem{DBLP:journals/entcs/GriesmayerSB07}
Griesmayer, A., Staber, S., Bloem, R.: Automated fault localization for {C} programs. In: Bloem, R., Roveri, M., Somenzi, F. (eds.) Proceedings of the Workshop on Verification and Debugging, V{\&}D@FLoC 2006, Seattle, WA, USA, August 21, 2006. Electronic Notes in Theoretical Computer Science, vol.~174, pp. 95--111. Elsevier (2006). \doi{10.1016/J.ENTCS.2006.12.032}, \url{https://doi.org/10.1016/j.entcs.2006.12.032}

\bibitem{iscas85}
Hansen, M.C., Yalcin, H., Hayes, J.P.: Unveiling the {ISCAS-85} benchmarks: {A} case study in reverse engineering. {IEEE} Des. Test Comput.  \textbf{16}(3),  72--80 (1999). \doi{10.1109/54.785838}, \url{https://doi.org/10.1109/54.785838}

\bibitem{imms18-PySAT}
Ignatiev, A., Morgado, A., Marques{-}Silva, J.: {PySAT}: {A} python toolkit for prototyping with {SAT} oracles. In: Beyersdorff, O., Wintersteiger, C.M. (eds.) Theory and Applications of Satisfiability Testing - {SAT} 2018 - 21st International Conference, {SAT} 2018, Held as Part of the Federated Logic Conference, FloC 2018, Oxford, UK, July 9-12, 2018, Proceedings. Lecture Notes in Computer Science, vol. 10929, pp. 428--437. Springer (2018). \doi{10.1007/978-3-319-94144-8\_26}, \url{https://doi.org/10.1007/978-3-319-94144-8\_26}

\bibitem{imms19-RC2}
Ignatiev, A., Morgado, A., Marques{-}Silva, J.: {RC2:} an efficient {MaxSAT} solver. J. Satisf. Boolean Model. Comput.  \textbf{11}(1),  53--64 (2019)

\bibitem{ijcai19-ignatievMWM}
Ignatiev, A., Morgado, A., Weissenbacher, G., Marques{-}Silva, J.: Model-based diagnosis with multiple observations. In: Kraus, S. (ed.) Proceedings of the Twenty-Eighth International Joint Conference on Artificial Intelligence, {IJCAI} 2019, Macao, China, August 10-16, 2019. pp. 1108--1115. ijcai.org (2019). \doi{10.24963/IJCAI.2019/155}, \url{https://doi.org/10.24963/ijcai.2019/155}

\bibitem{bugAssist-cav11}
Jose, M., Majumdar, R.: Bug-assist: Assisting fault localization in {ANSI-C} programs. In: Gopalakrishnan, G., Qadeer, S. (eds.) Computer Aided Verification - 23rd International Conference, {CAV} 2011, Snowbird, UT, USA, July 14-20, 2011. Proceedings. Lecture Notes in Computer Science, vol.~6806, pp. 504--509. Springer (2011). \doi{10.1007/978-3-642-22110-1\_40}, \url{https://doi.org/10.1007/978-3-642-22110-1\_40}

\bibitem{bugAssist-pldi11}
Jose, M., Majumdar, R.: Cause clue clauses: error localization using maximum satisfiability. In: Proceedings of the 32nd {ACM} {SIGPLAN} Conference on Programming Language Design and Implementation, {PLDI} 2011. pp. 437--446. {ACM} (2011)

\bibitem{DBLP:conf/fmcad/KonighoferB11}
K{\"{o}}nighofer, R., Bloem, R.: Automated error localization and correction for imperative programs. In: Bjesse, P., Slobodov{\'{a}}, A. (eds.) International Conference on Formal Methods in Computer-Aided Design, {FMCAD} '11, Austin, TX, USA, October 30 - November 02, 2011. pp. 91--100. {FMCAD} Inc. (2011), \url{http://dl.acm.org/citation.cfm?id=2157671}

\bibitem{DBLP:conf/icfem/LamraouiN14}
Lamraoui, S., Nakajima, S.: A formula-based approach for automatic fault localization of imperative programs. In: Merz, S., Pang, J. (eds.) Formal Methods and Software Engineering - 16th International Conference on Formal Engineering Methods, {ICFEM} 2014, Luxembourg, Luxembourg, November 3-5, 2014. Proceedings. Lecture Notes in Computer Science, vol.~8829, pp. 251--266. Springer (2014). \doi{10.1007/978-3-319-11737-9\_17}, \url{https://doi.org/10.1007/978-3-319-11737-9\_17}

\bibitem{jip16-SNIPER}
Lamraoui, S., Nakajima, S.: A formula-based approach for automatic fault localization of multi-fault programs. J. Inf. Process.  \textbf{24}(1),  88--98 (2016). \doi{10.2197/IPSJJIP.24.88}, \url{https://doi.org/10.2197/ipsjjip.24.88}

\bibitem{DBLP:journals/jar/LiffitonS08}
Liffiton, M.H., Sakallah, K.A.: Algorithms for computing minimal unsatisfiable subsets of constraints. J. Autom. Reason.  \textbf{40}(1),  1--33 (2008). \doi{10.1007/S10817-007-9084-Z}, \url{https://doi.org/10.1007/s10817-007-9084-z}

\bibitem{liu2019you}
Liu, K., Koyuncu, A., Bissyand{\'e}, T.F., Kim, D., Klein, J., Le~Traon, Y.: You cannot fix what you cannot find! an investigation of fault localization bias in benchmarking automated program repair systems. In: 2019 12th IEEE conference on software testing, validation and verification (ICST). pp. 102--113. IEEE (2019)

\bibitem{ijcai15-Marques-SilvaJI15}
Marques{-}Silva, J., Janota, M., Ignatiev, A., Morgado, A.: Efficient model based diagnosis with maximum satisfiability. In: Yang, Q., Wooldridge, M.J. (eds.) Proceedings of the Twenty-Fourth International Joint Conference on Artificial Intelligence, {IJCAI} 2015, Buenos Aires, Argentina, July 25-31, 2015. pp. 1966--1972. {AAAI} Press (2015), \url{http://ijcai.org/Abstract/15/279}

\bibitem{DBLP:journals/jair/MetodiSKC14}
Metodi, A., Stern, R., Kalech, M., Codish, M.: A novel sat-based approach to model based diagnosis. J. Artif. Intell. Res.  \textbf{51},  377--411 (2014). \doi{10.1613/JAIR.4503}, \url{https://doi.org/10.1613/jair.4503}

\bibitem{DBLP:journals/tosem/NaishLR11}
Naish, L., Lee, H.J., Ramamohanarao, K.: A model for spectra-based software diagnosis. {ACM} Trans. Softw. Eng. Methodol.  \textbf{20}(3),  11:1--11:32 (2011). \doi{10.1145/2000791.2000795}, \url{https://doi.org/10.1145/2000791.2000795}

\bibitem{C-Pack-IPAs}
Orvalho, P., Janota, M., Manquinho, V.: {C-Pack of IPAs: {A} {C90} Program Benchmark of Introductory Programming Assignments}. CoRR  \textbf{abs/2206.08768} (2022). \doi{10.48550/arXiv.2206.08768}, \url{https://doi.org/10.48550/arXiv.2206.08768}

\bibitem{InvAASTCluster-corr22}
Orvalho, P., Janota, M., Manquinho, V.: {InvAASTCluster: On Applying Invariant-Based Program Clustering to Introductory Programming Assignments}. CoRR  \textbf{abs/2206.14175} (2022). \doi{10.48550/ARXIV.2206.14175}, \url{https://doi.org/10.48550/arXiv.2206.14175}

\bibitem{fse22-MultIPAS}
Orvalho, P., Janota, M., Manquinho, V.: Mult{IPA}s: {A}pplying {P}rogram {T}ransformations to {I}ntroductory {P}rogramming {A}ssignments for {D}ata {A}ugmentation. In: Proceedings of the 30th {ACM} Joint European Software Engineering Conference and Symposium on the Foundations of Software Engineering, {ESEC/FSE} 2022. pp. 1657--1661. {ACM}, Singapore (2022). \doi{10.1145/3540250.3558931}

\bibitem{C-Pack-IPAs_apr24}
Orvalho, P., Janota, M., Manquinho, V.: {C-Pack of IPAs: {A} {C90} Program Benchmark of Introductory Programming Assignments}. In: International Workshop on Automated Program Repair, APR@ICSE 2024, Lisbon, Portugal, April 20, 2024. pp.~-- (2024). \doi{10.1145/3643788.3648010}, \url{https://doi.org/10.1145/3643788.3648010}

\bibitem{CFaults-Zenodo-FM2024}
Orvalho, P., Janota, M., Manquinho, V.: {CFaults: Model-Based Diagnosis for Fault Localization in C with Multiple Test Cases} (Jun 2024). \doi{10.5281/zenodo.12510220}, \url{https://github.com/pmorvalho/CFaults}

\bibitem{ecai23-GNNs-4-var-mapping}
Orvalho, P., Piepenbrock, J., Janota, M., Manquinho, V.M.: Graph neural networks for mapping variables between programs. In: {ECAI} 2023 - 26th European Conference on Artificial Intelligence. Frontiers in Artificial Intelligence and Applications, vol.~372, pp. 1811--1818. {IOS} Press, Poland (2023). \doi{10.3233/FAIA230468}, \url{https://doi.org/10.3233/FAIA230468}

\bibitem{pycparser}
pycparser: {}. \url{https://github.com/eliben/pycparser} (2024), [Online; accessed 18-April-2024]

\bibitem{reiter87}
Reiter, R.: A theory of diagnosis from first principles. Artif. Intell.  \textbf{32}(1),  57--95 (1987). \doi{10.1016/0004-3702(87)90062-2}, \url{https://doi.org/10.1016/0004-3702(87)90062-2}

\bibitem{DBLP:conf/cav/RothenbergG20}
Rothenberg, B., Grumberg, O.: Must fault localization for program repair. In: Lahiri, S.K., Wang, C. (eds.) Computer Aided Verification - 32nd International Conference, {CAV} 2020, Los Angeles, CA, USA, July 21-24, 2020, Proceedings, Part {II}. Lecture Notes in Computer Science, vol. 12225, pp. 658--680. Springer (2020). \doi{10.1007/978-3-030-53291-8\_33}, \url{https://doi.org/10.1007/978-3-030-53291-8\_33}

\bibitem{DBLP:conf/fmcad/SafarpourMVLS07}
Safarpour, S., Mangassarian, H., Veneris, A.G., Liffiton, M.H., Sakallah, K.A.: Improved design debugging using maximum satisfiability. In: Formal Methods in Computer-Aided Design, 7th International Conference, {FMCAD} 2007, Austin, Texas, USA, November 11-14, 2007, Proceedings. pp. 13--19. {IEEE} Computer Society (2007). \doi{10.1109/FAMCAD.2007.26}, \url{https://doi.org/10.1109/FAMCAD.2007.26}

\bibitem{DBLP:journals/ese/SoremekunKBZ21}
Soremekun, E.O., Kirschner, L., B{\"{o}}hme, M., Zeller, A.: Locating faults with program slicing: an empirical analysis. Empir. Softw. Eng.  \textbf{26}(3), ~51 (2021). \doi{10.1007/S10664-020-09931-7}, \url{https://doi.org/10.1007/s10664-020-09931-7}

\bibitem{DBLP:journals/jss/WongDC10}
Wong, W.E., Debroy, V., Choi, B.: A family of code coverage-based heuristics for effective fault localization. J. Syst. Softw.  \textbf{83}(2),  188--208 (2010). \doi{10.1016/J.JSS.2009.09.037}, \url{https://doi.org/10.1016/j.jss.2009.09.037}

\bibitem{DBLP:journals/tr/WongDGL14}
Wong, W.E., Debroy, V., Gao, R., Li, Y.: The dstar method for effective software fault localization. {IEEE} Trans. Reliab.  \textbf{63}(1),  290--308 (2014). \doi{10.1109/TR.2013.2285319}, \url{https://doi.org/10.1109/TR.2013.2285319}

\bibitem{DBLP:journals/tse/WongGLAW16}
Wong, W.E., Gao, R., Li, Y., Abreu, R., Wotawa, F.: A survey on software fault localization. {IEEE} Trans. Software Eng.  \textbf{42}(8),  707--740 (2016). \doi{10.1109/TSE.2016.2521368}, \url{https://doi.org/10.1109/TSE.2016.2521368}

\bibitem{DBLP:journals/jlp/WotawaNM12}
Wotawa, F., Nica, M., Moraru, I.: Automated debugging based on a constraint model of the program and a test case. J. Log. Algebraic Methods Program.  \textbf{81}(4),  390--407 (2012). \doi{10.1016/J.JLAP.2012.03.002}, \url{https://doi.org/10.1016/j.jlap.2012.03.002}

\bibitem{DBLP:conf/popl/XieA05}
Xie, Y., Aiken, A.: Scalable error detection using boolean satisfiability. In: Palsberg, J., Abadi, M. (eds.) Proceedings of the 32nd {ACM} {SIGPLAN-SIGACT} Symposium on Principles of Programming Languages, {POPL} 2005, Long Beach, California, USA, January 12-14, 2005. pp. 351--363. {ACM} (2005). \doi{10.1145/1040305.1040334}, \url{https://doi.org/10.1145/1040305.1040334}

\bibitem{zeller1999yesterday}
Zeller, A.: Yesterday, my program worked. today, it does not. why? In: ESEC/FSE'99, 7th European Software Engineering Conference, Held Jointly with the 7th {ACM} {SIGSOFT} Symposium on the Foundations of Software Engineering 1999. Lecture Notes in Computer Science, vol.~1687, pp. 253--267. Springer (1999)

\end{thebibliography}
\end{document}